\documentclass[conference]{IEEEtran}
\IEEEoverridecommandlockouts
\usepackage{cite}
\usepackage{amsmath,amssymb,amsfonts}
\usepackage{algorithmic}
\usepackage{graphicx}
\usepackage{textcomp}
\usepackage{xcolor}

\newcommand{\mx}{\ensuremath{_{\text{max}}}}
\newcommand{\mn}{\ensuremath{_{\text{min}}}}
\newcommand{\mob}{\ensuremath{_{\text{mob}}}}

\usepackage{dsfont}
\usepackage{subcaption} 
\usepackage{soul}

\definecolor{warninggreen}{RGB}{70, 161, 4}       

\usepackage{subcaption} 
\usepackage{bbm} 
\usepackage{mathtools} 
\DeclareMathOperator*{\argmax}{arg\,max} 

\def\BibTeX{{\rm B\kern-.05em{\sc i\kern-.025em b}\kern-.08em
    T\kern-.1667em\lower.7ex\hbox{E}\kern-.125emX}}
\begin{document}

\title{Cruising the Spectrum: Joint Spectrum Mobility and Antenna Array Management for Mobile (cm/mm)Wave Connectivity
\\
\thanks{This work has been supported in part by NSF through grants 2030141, 2112471, CNS 2030272 and CNS 2212050.}
}

\author{\IEEEauthorblockN{Ece Bing\"{o}l, Eylem Ekici}
\IEEEauthorblockA{\textit{Department of Electrical and Computer Engineering} \\
\textit{The Ohio State University}\\
Columbus, OH \\
\{bingol.2, ekici.2\} @osu.edu}
\and
\IEEEauthorblockN{Mehmet C. Vuran}
\IEEEauthorblockA{\textit{School of Computing} \\
\textit{University of Nebraska - Lincoln}\\
Lincoln, NE \\
mcv@unl.edu}
}

\maketitle

\begin{abstract}

The large bandwidths available at millimeter wave (mmWave) FR2 bands (24--71 GHz) and the emerging FR3 bands (7--24 GHz) are essential for supporting high data rates. Highly directional beams utilized to overcome the attenuation in these frequencies necessitate robust and efficient beamforming schemes. Nevertheless, antenna and beam management approaches still face challenges in highly mobile solutions, such as vehicular connectivity, with increasing number of bands. In this work, the concept of spectrum mobility is studied along with antenna array management in multiple frequencies to improve beamforming under mobility. The spectrum mobility problem aims to select the optimal channel frequency and beam direction in each time slot to maximize data rate. This problem is formulated as a Partially Observable Markov Decision Process (POMDP), and Point-Based Value Iteration (PBVI) algorithm is used to find a policy with performance guarantees.
Numerical examples confirm the efficacy of the resulting policy for multiple available frequency bands, even when the user mobility significantly deviates from models assumed during policy generation. 

\end{abstract}


\section{Introduction}

The advent of 5G wireless networks gave rise to various high-throughput applications such as Virtual Reality (VR), Internet of Things (IoT), Unmanned Aerial Vehicles (UAV), Industry 4.0 and Cellular Vehicle-to-Everything (C-V2X) networks~\cite{erunkulu-5g-use-cases-survey}. In return, traffic demand is expected to increase exponentially~\cite{giordani-6g-survey}. This places a significant strain on the spectral resources. To this end, FR2 bands (24--71 GHz) in the mmWave region are allocated to 5G~\cite{Dilli-FR1-FR2-analysis}, and FR3 cmWave bands (7--24 GHz) have also been recently proposed for deployment~\cite{rappaport-fr1-fr3-propagation-models}. Large bandwidths available at these frequencies can support high data rates. However, link maintenance becomes an important issue, especially in mmWave systems. Severe attenuation experienced in these frequency bands requires the use of highly focused beams that are generated using antenna arrays. Therefore, beamforming and tracking are essential features of 5G systems.

Numerous beamforming and beam tracking techniques have been studied in the literature~\cite{Xue-beamforming-survey}. 
Extended Kalman Filter (EKF)-based algorithms are proposed in~\cite{zhang_AoD_tracking_EKF} and~\cite{heath-analog-beamtracking-ekf} to track the angle of departure (AoD) and angle of arrival (AoA). Motivated by monopulse radar systems, an auxiliary beam pair (ABP) design for AoA/AoD estimation is developed in~\cite{dalin-aux-beam-pair-mmwave}. In~\cite{kim-abp-q-learning-mmwave}, ABP is combined with Q-learning to obtain a model-free beam tracking algorithm. Beam discovery problem is presented as an error detection problem in~\cite{ekici_Beam_disc_with_lin_block_codes} and a beam discovery method inspired by linear block codes is proposed. An adaptive beamwidth control algorithm is developed in~\cite{kim_adaptive_beamwidth_control} for robust beamtracking. In~\cite{seo-mmwave-pomdp}, Partially Observable Markov Decision Process (POMDP) framework is used to formulate the beamforming problem. The operating frequency is considered to be fixed throughout these works. Meanwhile, in~\cite{heath_rainbow}, True Time Delay (TTD) elements are employed to generate frequency-dependent rainbow beams. These beams are used to track relay vehicles with a small time overhead. A flexible directional-frequency multiplexing system is developed in~\cite{jain-mmFlexible}, by combining delay and phase elements. The proposed system can modify the frequency bands assigned to users regardless of their direction. Beamforming in underground wireless networks is studied in~\cite{vuran-smart-ug-arrays} and it is shown that the wavelength changes with soil moisture. Hence, an adaptive beamforming algorithm is designed to overcome the resulting adversaries.

A common thread in these and other work in the literature is the use of a single frequency range for the communication system. However, advances in antenna and front-end design practices allow the use of the same transceiver in \textit{multiple frequency ranges}. Access to multiple frequency bands definitely increases the access to available bandwidth, along with advantages garnered from statistical multiplexing. We argue that the advantages of multi spectrum access go further: The propagation characteristics in different bands, combined with the selections of the antenna, bandwidth, and beam form vectors, create opportunities to address various mobility-related problems. The interdependencies in signal resource allocation decisions lead to non-trivial optimization problems. As an example, higher frequency ranges usually have larger channel bandwidth, but require a larger number of antennas to form narrower beams to achieve a target SNR. However, narrow beams require frequent updates to counter end node mobility. On the other hand, lower frequency bands can achieve a target SNR level with smaller number of antennas and wider beams. While this is advantageous in handling mobility, these bands usually have smaller channel bandwidth. These tradeoffs are further complicated by variation in distances, uncertainties in end node mobility, and channel availability.

In this work, we investigate the problem of managing the spectrum and antenna arrays under end node mobility, with the express intent of revealing advantages of using resources over a wide range of the EM spectrum. More specifically, we consider a base station (BS) and mobile user equipments (UEs), and make resource allocation decisions in terms of frequency selection and beam management. BS's resource allocation decisions are based on its observations of the communication channel with each individual UE and statistical mobility models: BS \textit{cruises the spectrum} over time to find the best data rate for individual UEs at that moment. The problem is formulated as a Partially Observable Markov Decision Process (POMDP) and solved using a Point-Based Value Iteration (PBVI) algorithm. Given a stochastic mobility model, the proposed algorithm provides an approximation of the optimal policy with bounded error.

The rest of the paper is organized as follows: Related work is discussed in Section~\ref{sec:related-work}. System model is explained in Section~\ref{sec:sys-model}. Solution approach and numerical results are presented in Sections~\ref{sec:solution} and~\ref{sec:numeric}, respectively. Finally, the concluding remarks and future directions are discussed in Section~\ref{sec:conclusions}.

\section{Related Work} \label{sec:related-work}

Most of the literature focuses on beamforming under fixed channel frequency and array architecture~\cite{Xue-beamforming-survey}. However, as pointed out in works such as~\cite{kim_adaptive_beamwidth_control,vuran-smart-ug-arrays,heath_rainbow,poor_delay_phase_precoding_thz,jain-mmFlexible}, the efficiency and robustness of beamforming procedures can be further improved by leveraging the frequency and array element dependencies in beam patterns.

In~\cite{kim_adaptive_beamwidth_control}, adjusting the beamwidth by partial activation of the antenna array is proposed to increase the robustness of beamtracking. The main goal is to mitigate the drop in Signal-to-Noise Ratio (SNR) due to beam misalignment resulting from mobility. The proposed algorithm is shown to achieve higher SNR levels and lower AoA estimation error compared to beamtracking without beamwidth control. The objective of~\cite{kim_adaptive_beamwidth_control} is similar to ours because, for a given bandwidth, the data rate increases with SNR. However, since the primary focus is on the array management, the operating frequency is considered to be fixed in~\cite{kim_adaptive_beamwidth_control}, where spectrum mobility is the primary focus of our work.

Underground wireless communications has been studied in~\cite{vuran-smart-ug-arrays}, where the wavelength changes as a function of soil moisture. An adaptive beamforming approach has been proposed to maintain high directivity and desired beam shape. 
Array thinning, which is the activation of particular subsets of the array elements, is used to reduce the sidelobes. 
While changes in communication frequency are considered in~\cite{vuran-smart-ug-arrays}, these changes are due to environmental conditions. 
In contrast, we consider frequency as a resource and a decision variable, aiming to maximize the expected data rate under mobility. 

Frequency-dependent rainbow beams are studied in~\cite{poor_delay_phase_precoding_thz,heath_rainbow}. The beam split effect where different Orthogonal Frequency Division Multiplexing (OFDM) subcarriers point at separate physical directions is investigated in~\cite{poor_delay_phase_precoding_thz} and a Delay Phase Precoding (DPP) architecture is developed to prevent this. On the other hand,~\cite{heath_rainbow} introduces beam split to track vehicles in relay networks. Both works consider OFDM systems within a single contiguous band. In contrast, we consider separate frequency bands with different propagation characteristics is is projected for 6G systems.

A frequency-dependent beamforming architecture enabling flexible directional-frequency multiplexing is developed in~\cite{jain-mmFlexible}. The system incorporates time delay and phase-shift elements together such that frequency resources can be allocated to users in any direction according to their demands. However, exploiting spectral resources to improve beamforming under mobility is beyond the scope of~\cite{jain-mmFlexible}.

\section{System Model} \label{sec:sys-model}

\subsection{Notation}
Boldface letters $\mathbf{x}$ and $\mathbf{X}$ denote vectors and matrices, respectively. $\mathbf{x}[i]$ denotes the $i^{th}$ element of vector $\mathbf{x}$. The transpose and Hermitian transpose of $\mathbf{X}$ are given by $\mathbf{X}^T$ and $\mathbf{X}^H$, respectively. $\mathbf{X} \otimes \mathbf{Y}$ is the Kronecker product of $\mathbf{X}$ and $\mathbf{Y}$. Sets are designated with $\mathcal{X}$ where $|\mathcal{X}|$ is cardinality. $\mathbb{N}$, $\mathbb{R}$ and $\mathbb{C}$  correspond to the sets of natural, real and complex numbers, respectively. The set of non-negative real numbers is denoted by $\mathbb{R}_{\geq 0}$. The set of integers from $1$ to $n$ is indicated by $\mathbb{N}_{\leq n}$, i.e. $\mathbb{N}_{\leq n} \coloneq \{1, \ldots, n\}$.  Complex Gaussian distribution with mean $\mu$ and variance $\sigma^2$ is denoted by $\mathcal{CN}(\mu, \sigma^2)$. Lastly, $\mathds{1}(\cdot)$ is the indicator function.

\subsection{Channel Model}

We consider a base station (BS) and a mobile user positioned on a grid. The user has an omnidirectional antenna with unity gain, whereas the BS is equipped with a uniform planar array (UPA). Time is slotted $t \in \{0,1,\ldots\}$ and the BS performs analog beamforming at the beginning of each time slot.

We assume that there are $C$ channels at the disposal of the BS. Each channel $c \in \{1,2, \ldots, C\}$ has a bandwidth, $W_c$, with center frequency, $f_c$. $W_c$'s are fixed but can be different for different channels. The set of channel frequencies is denoted by $\mathcal{F} = \{f_i\}_{i=1}^{C}$.

\begin{figure}[!t]
    \centering    \includegraphics[width=0.95\linewidth]{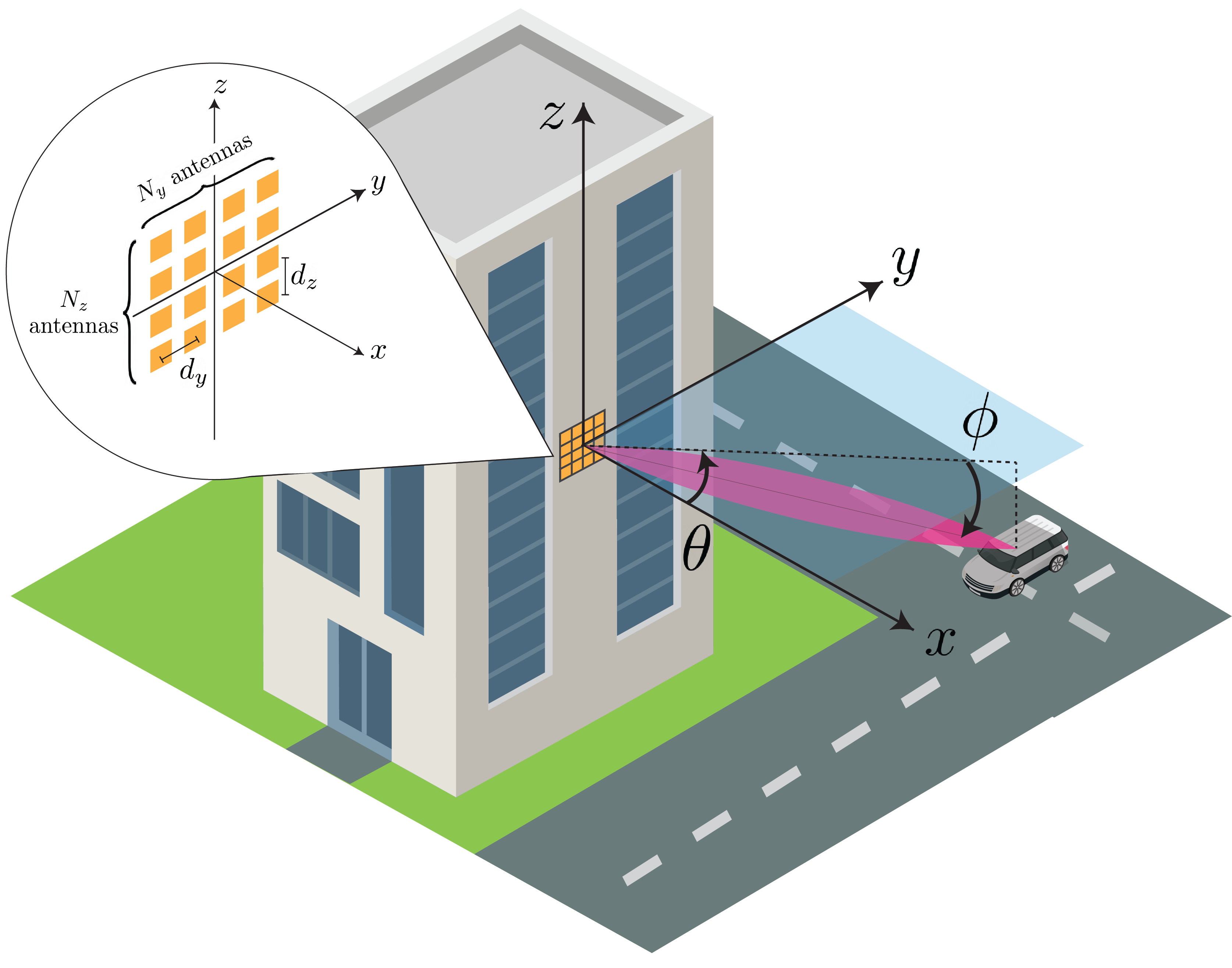}
    \caption{UPA in the 3D coordinate system}
    \label{fig:UPA}
\end{figure}

To support multiband operation, we assume that the BS is equipped with a shared-aperture antenna array~\cite{askarian-shared-aperture-antenna-review}. 
Instead of using separate antenna architectures for different frequency bands, recently, shared-aperture antennas are proposed to integrate multiple antennas into the same structure in which all sub-antennas radiate through a common surface~\cite{askarian-shared-aperture-antenna-review}. This also increases space efficiency, which might be a concern for mobile devices. Shared-aperture antennas have garnered significant interest over the past decade and various designs were proposed for applications in 5G/6G communications~\cite{ding-dual-band-shared-aperture-antenna-mmwave,mei-triple-band-shared-aperture-antenna,askarian-shared-aperture-antenna-review}. Motivated by this, we adopt a similar antenna architecture in this paper. We assume that each channel has its dedicated UPA and these UPAs are integrated into a common radiating surface of area $A_y \times A_z$. One of such UPAs is shown in Fig.~\ref{fig:UPA}. Each rectangle represents an antenna connected to a phase shifter. The array is positioned on the $yz$-plane. It consists of $N_y \times N_z$ antennas in total, where $N_y$ and $N_z$ are the number antennas on $y$ and $z$ directions, respectively. Distances between adjacent antennas in $y$ and $z$ directions are given by $d_y$ and $d_z$, respectively. Azimuth angle is denoted by $\theta$ and is measured counterclockwise from the $x$-axis. Similarly, elevation angle is indicated by $\phi$ and is measured upwards (towards $z$-axis) from the $xy$-plane. We consider $\theta, \phi \in [-\pi/2, \pi/2]$ in this work. Suppose that the UPA of channel $c \in \{1,2, \ldots, C\}$ consists of $N_y^c \times N_z^c$ antennas and the inter-element spacings are $d_y^c$ and $d_z^c$, respectively. Then, the array factor can be expressed as~\cite{balanis-antenna-theory}:
\begin{equation} \label{eq:array_factor}
    \begin{aligned}
        AF_c(\theta, \phi) = \sum_{m=1}^{N_y^c} \sum_{n=1}^{N_z^c} \exp \Biggl(j \biggl[ & (m-1) \frac{2\pi}{\lambda_c}  d_y^c \cos \phi \sin\theta \\
        +  &(n-1) \frac{2\pi}{\lambda_c}  d_z^c \sin\phi + \delta^c_{mn} \biggr] \Biggr),
    \end{aligned}
\end{equation}
where $\lambda_c$ is the wavelength associated with frequency $f_c$, $\delta_{mn}^c$ is the phase shift applied to the $mn^{th}$ antenna element and $(\theta, \phi)$ is the azimuth and elevation angle pair measured from the antenna boresight as shown in Fig.~\ref{fig:UPA}. Phase shifts required to point the beam in a specific direction $(\hat{\theta}, \hat{\phi})$ are~\cite{haupt-timed-arrays}:
\begin{equation} \label{eq:phase_shifts}
    \delta_{mn}^c = - \frac{2 \pi}{\lambda_c} \left[ (m-1) d_y^c \cos \hat{\phi} \sin \hat{\theta} + (n-1) d_z^c \sin \hat{\phi} \right] \forall m,n.
\end{equation}
Note that, $N_y^c$ and $N_z^c$ depend on the aperture area $A_y \times A_z$ and inter-element spacings $d_y^c$ and $d_z^c$. Intuitively, as $f_c$ increases, more antennas can be packed into the same area because the antenna element spacing is proportional to the wavelength. Throughout this paper, we assume that the elements of each UPA are critically separated with respect to its intended operating frequency $f_c$, i.e., $d_y^c = d_z^c = 0.5 \lambda_c$. Then, t array steering vector of the UPA is expressed as~\cite{chen-near-far-field-channel-model-mimo}: 
\begin{equation} \label{eq:3D_steering_vec}
    \mathbf{a}_c (\theta, \phi) = \mathbf{a}_y^c (\theta, \phi) \otimes \mathbf{a}_z^c (\phi) \ \in \mathbb{C}^{N_y^c N_z^c \times 1}.
\end{equation}
$\mathbf{a}_y^c (\theta, \phi)$ and $\mathbf{a}_z^c (\phi)$ are the array steering vectors in horizontal (azimuth) and vertical (elevation) directions, given by~\cite{chen-near-far-field-channel-model-mimo}:
\begin{subequations} \label{eq:2D_steering_vecs}
\begin{align}
    \mathbf{a}_y^c (\theta, \phi) &= 
    \begin{bmatrix}
            1, e^{-j 2\pi \psi_c}, \ldots,  e^{-j 2\pi (N_y^c-1) \psi_c}
        \end{bmatrix}^T \in \mathbb{C}^{N_y^c \times 1}, \label{eq:hori_steering_vec} \\
    \mathbf{a}_z^c (\phi) &=
    \begin{bmatrix}
            1, e^{-j 2\pi \zeta_c}, \ldots,  e^{-j 2\pi (N_z^c-1) \zeta_c}
        \end{bmatrix}^T \in \mathbb{C}^{N_z^c \times 1},
    \label{eq:vert_steering_vec}
\end{align}
\end{subequations}
where $\psi_c$ and $\zeta_c$ are the normalized spatial angles related to the physical angles $\theta, \phi \in [-\pi/2, \pi/2]$ as
\begin{equation} \label{eq:norm_spatial_angs}
    \psi_c = \frac{d_y^c \cos \phi \sin \theta}{\lambda_c} \quad\mathrm{and}\quad \zeta_c = \frac{d_z^c \sin \phi}{\lambda_c}. 
\end{equation}
Using the steering vectors in (\ref{eq:3D_steering_vec})-(\ref{eq:norm_spatial_angs}), the channel vector of frequency band $f_c \in \mathcal{F}$ at time slot $t$ can be expressed as
\begin{equation}\label{eq:ch_vec_multipath}
    \mathbf{h}_t^{c} = \sum_{l=1}^{L} \beta_{t}^{l,c} \ \mathbf{a}_c^{H}(\theta_t^{l}, \phi_t^{l}) \quad \forall c\in \{1, \ldots, C\},
\end{equation}
where $L$ is the number of paths, $\beta_t^{l,c}$ is the complex channel gain and $(\theta_t^{l}, \phi_t^{l})$ is the and Angle of Departure (AoD) pair of the $l^{th}$ path in time slot $t$~\cite{heath_overview_of_SP_techniques_for_mmwave_mimo}. In this work, we concentrate on the Line-of-Sight (LoS) path and ignore the fading effects. Thus, the path index $l$ is dropped in the rest of the paper for simplicity. Furthermore, we use path gain for $\beta_t^{c}$, i.e.,
\begin{equation}\label{eq:path_loss}
    \beta_t^{c} = \sqrt{\frac{K}{r_t^{\eta} f_c^2 }} \ ,
\end{equation}
where $r_t$ is the distance between the BS and the user in time slot $t$, $\eta$ is the path loss exponent and $K$ is a constant. The implication of (\ref{eq:path_loss}) is that the gain decreases with frequency. If channel frequencies $f_c \in \mathcal{F}$ are close, this effect might be negligible. However, if the center frequencies are significantly separate, channel gains could differ substantially. Channels with higher center frequencies experience more attenuation. This is compensated by antenna array size. Hence, the number of antennas in a UPA increases with frequency since more elements can be accommodated in a given area. Under these assumptions, the channel vector of frequency band $f_c$ at time $t$ becomes
\begin{equation} \label{eq:ch_vec_simplified}
    \mathbf{h}_t^{c} = \sqrt{\frac{K}{r_t ^ {\eta} f_c^2}} \ \mathbf{a}_c^{H} \big(\theta_t, \phi_t \big) \quad \forall c\in \{1, \ldots, C\}.
\end{equation}

Suppose that the BS selects a channel $q(t) \in \mathbb{N}_{\leq C}$ with center frequency $f_{q(t)} \in \mathcal{F}$ and forms a beam in direction $(\hat{\theta}_t, \hat{\phi}_t)$ in time slot $t$. The message signal $x_t$ is first passed through a splitter then multiplied by the analog combiner vector in (\ref{eq:3D_steering_vec}). The resulting baseband signal at the array output is given by
\begin{equation} \label{eq:trans_sig}
    \mathbf{\tilde{x}}_t = \sqrt{\frac{P_T}{N_y^{q(t)} N_z^{q(t)}}} \ \mathbf{a}_{q(t)} ( \hat{\theta}_t, \hat{\phi}_t) x_t,
\end{equation}
where $P_T$ is the total transmit power, i.e., $\mathbf{\tilde{x}}_t^H \mathbf{\tilde{x}}_t = P_T$. It is fixed and distributed equally between all $N_y^{q(t)} \times N_z^{q(t)}$ antennas. Signal $x_t$ is considered to have unit power.
 
Let the actual position of the user at time $t$ with respect to the BS be given by the spherical coordinates $(r_t, \theta_t, \phi_t)$. Using the channel response in (\ref{eq:ch_vec_simplified}), the received signal in time slot $t$ can be expressed as
\begin{small}
\begin{equation} \label{eq:rec_sig}
    \begin{aligned}
        y_t &= \mathbf{h}_t^{q(t)} \mathbf{\tilde{x}}_t + n^{q(t)}_t \\
        &= \sqrt{\frac{K P_T}{r_t ^ {\eta} f_{q(t)}^2 N_y^{q(t)} N_z^{q(t)}} } \ \mathbf{a}_{q(t)}^{H}(\theta_t, \phi_t) \ \mathbf{a}_{q(t)}(\hat{\theta}_t, \hat{\phi}_t) x_t + n^{q(t)}_t,
    \end{aligned}
\end{equation}\end{small}
where $n^{q(t)}_t$ is the complex Gaussian noise with zero mean and variance $\sigma_{q(t)}^2$, i.e., $n^{q(t)}_t \sim \mathcal{CN}(0, \sigma_{q(t)}^2)$. Note that the noise variance depends on the channel index, because for white noise, the noise floor increases with bandwidth~\cite{proakis-comm-systems}. We assume that for a given channel, noise samples taken in different time slots are independent and identically distributed (i.i.d.).

Using (\ref{eq:rec_sig}), the signal-to-noise (SNR) ratio, $\gamma_t$ at time $t$ is calculated as
\begin{equation} \label{eq:SNR}
    \begin{aligned}
        \gamma_t &= \frac{(\mathbf{h}_t^{q(t)} \mathbf{\tilde{x}}_t) (\mathbf{h}_t^{q(t)} \mathbf{\tilde{x}}_t)^*}{\sigma_{q(t)}^2} \quad \in \mathbb{R}_{\geq 0}, \\
    \end{aligned}
\end{equation}
and the corresponding maximum achievable rate is given by
\begin{equation} \label{eq:rate}
    R_t = W_{q(t)} \log_2{(1+ \gamma_t)} \quad \in \mathbb{R}_{\geq 0}.
\end{equation}

\begin{figure}[!t]
    \centering    \includegraphics[width=0.99\linewidth]{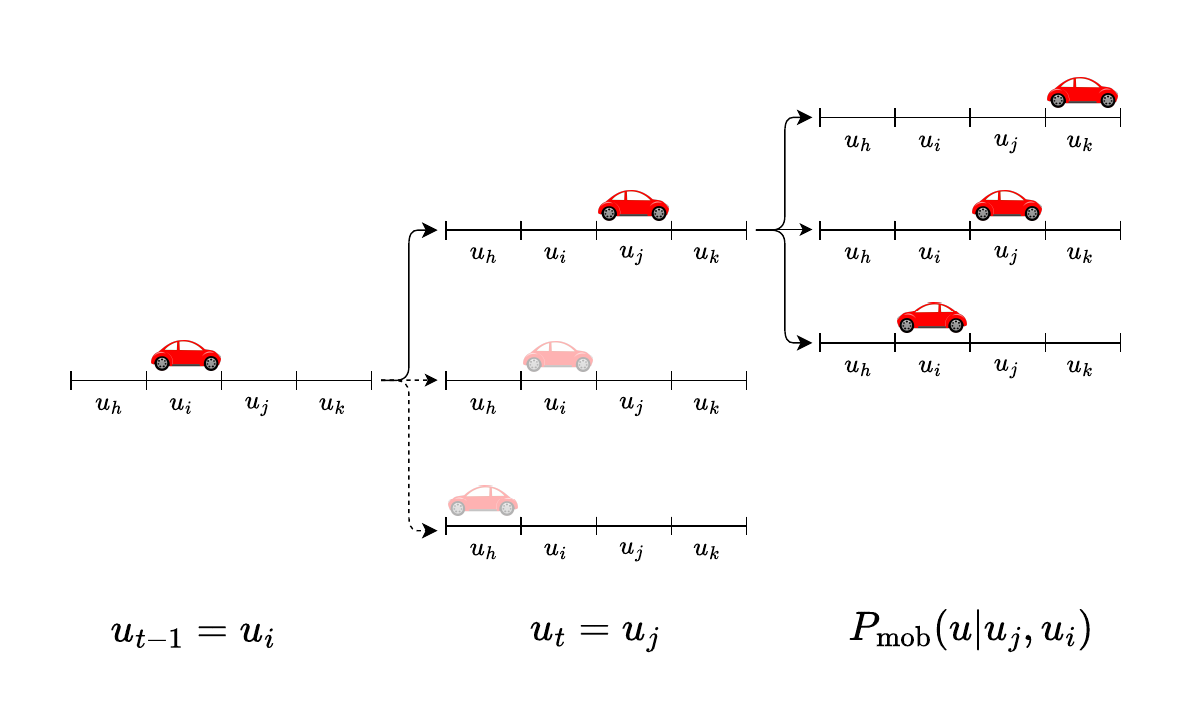}
    \caption{Position of the user according to the mobility model}
    \label{fig:mobility_model}
\end{figure}

\subsection{Mobility Model}
We impose a grid structure on the cell. The BS is stationary and located at the origin, whereas the user travels along a road on the grid. The road is represented by a set of grid cells $\mathcal{U} = \{u_1, \dots, u_{M_r} \}$. The cardinality of this set is $|\mathcal{U}| = M_r$. Each cell $u_i \in \mathcal{U}$ is represented by the spherical coordinates $(r^{u_i}, \theta^{u_i}, \phi^{u_i})$ of its center. The set of all azimuth and elevation angles are denoted with $\Theta = \{\theta^{u_i} \}_{i=1}^{M_r}$ and $\Phi = \{\phi^{u_i}\}_{i=1}^{M_r}$, respectively. 

The motion of the user along the road is statistically represented by a mobility model. Given the position of the user in the last $w$ slots, the model gives the probability of the user being in cell $u_i \in \mathcal{U}$ in the next time slot. This is illustrated in Fig.~\ref{fig:mobility_model} for $w=2$.  We assume that the mobility model is stationary, i.e., does not depend on time. Let the position of the user in a window of size $w$ be $\mathbf{u}_t^w = [u_{t-w+1},\ldots,u_{t}]^T$. Then,
\begin{equation}\label{eq:mob_model}
    \begin{aligned}
        P\mob(u_i, \mathbf{u}_{t-1}^w) &\coloneq P(u_t = u_i | u_{t-1}, \ldots, u_{t-w}), \quad u_i \in \mathcal{U} \\
        &= P(u_{w+1} = u_i | u_{w}, \ldots, u_{1}) \quad u_i \in \mathcal{U}.
    \end{aligned}
\end{equation}
The second equality comes from the stationarity assumption. We assume that the spherical coordinates $(r^{u_i}, \theta^{u_i}, \phi^{u_i})$ $ \forall u_i \in \mathcal{U}$ and the mobility model $P\mob(\cdot, \cdot)$ are available at the BS.



\subsection{Problem Statement}

At each time slot $t$, the BS selects a channel $q(t) \in \{1, \ldots, C\}$ with center frequency $f_{q(t)} \in \mathcal{F}$ and a beam direction $(\hat{\theta}_t, \hat{\phi}_t) \in \Theta \times \Phi$, then observes the resulting SNR, $\gamma_t$. The set of all decisions and observations until time $t$ is called history and denoted by $\mathcal{H}_t = \left\{\hat{\theta}_k, \hat{\phi}_k, q(k), f_{q(k)}, N_y^{q(k)}, N_z^{q(k)}, W_{q(k)}, \gamma_k \right\}_{k=1}^{t}$.

Ultimately, goal of the BS is to maximize the transfer of data to the user while satisfying the Quality of Service (QoS) constraints, such as minimum SNR requirements. However, beamforming under uncertainty imposes a unique challenge. As stated in (\ref{eq:rate}), bandwidth is a key factor in the rate. When the SNR level is fixed, the rate increases linearly with bandwidth, and generally more spectrum is available at higher frequencies. Hence, $W_{c}$ usually increases with $f_c$, which translates to higher data rates at higher frequencies under the same SNR. On the other hand, high frequencies experience more attenuation. As stated in (\ref{eq:path_loss}), for a given distance, path gain decreases with frequency, causing the SNR to drop. To counteract this, highly focused pencil beams are generated using antenna arrays. Arrays of higher frequencies usually consist of more antennas because more elements can be packed into the same area. This results in narrower beams, which alleviate the severe attenuation in higher frequencies. However, as the beams get narrower, any mismatch between the beam direction and the actual position of the user further degrades the performance. Narrow beams focus most of the transmitted power in a specific direction. Thus, the received power quickly drops as the user deviates from that direction. Normally, if the exact angular position $(\theta_t, \phi_t)$ of the user were known, the expected instantaneous rate would be maximized by steering the beam in that direction. For any given $(f_c, N_y^c, N_z^c, W_c)$ tuple, the maximum of (\ref{eq:rate}) occurs when $(\hat{\theta}_t, \hat{\phi}_t) = (\theta_t, \phi_t)$. Then, given the actual user position $(r_t, \theta_t, \phi_t)$, the optimal beam direction $(\hat{\theta}^*_t, \hat{\phi}^*_t)$ and the channel index $q^*(t)$ become
\begin{subequations} \label{eq:opt_sol_perf_info}
\begin{align}
    \hat{\theta}_t^* &= \theta_t \quad\mathrm{and}\quad  \hat{\phi}^*_t = \phi_t, \label{eq:opt_dir_perf_info} \\
    q^*(t) &=\argmax_{c \in \{1, \ldots, C\}} \ E\left[W_c \log_2 \left(1+ \frac{K P_T N_y^c N_z^c}{r_t^{\eta} f_c^2 |n_t^{c}|^2} \right) \right]. \label{eq:opt_freq_perf_info}
\end{align}
\end{subequations}
The expectation is with respect to the noise. It is implied in (\ref{eq:opt_freq_perf_info}) that when $(r_t, \theta_t, \phi_t)$ is known, the optimal channel index $q^*(t)$ is fixed and independent of time. Nonetheless, under mobility, the actual position of the user is usually not known and must be estimated by the BS. When the uncertainty in the estimation is high, the BS can cast wide beams using low frequencies to cover large areas. This increases the probability of having the user within the half-power beamwidth. However, if the estimation error is small, the BS loses the opportunity to transmit data at a higher rate using high frequencies. In addition, assuming the user is covered by the beam, uncertainty in the true position of the user persists, since it might be anywhere within the beamwidth. In contrast, BS can exploit high frequency channels to achieve high data rates when the uncertainty is low. This also helps to keep the uncertainty low, as long as the user is within the beamwidth. Since the beams generated at high frequencies are narrow, it is easier to pinpoint the actual location of the user within the beamwidth. However, if the estimation error is large enough, the user might not be covered by the beam. This significantly degrades performance. Moreover, once the tracking failure occurs, the uncertainty increases greatly, and the BS has to locate the user again. Therefore, BS has to balance risk and reward when making decisions under uncertainty.

To capture the trade-offs between frequency, bandwidth and array size under mobility we formulate the following maximization problem in (\ref{eq:obj_func}). We define policy $\pi(\cdot)$ as a mapping from history to a beam direction and channel index pair, i.e., $\pi(\cdot): \mathcal{H}_{t-1} \rightarrow \left( (\hat{\theta}_t , \hat{\phi}_t), q(t) \right),$ $ \forall t$ and denote the set of policies by $\Pi$. Our objective is to find a policy $\pi \in \Pi$ such that the expectation of the discounted sum rate is maximized. 
\begin{equation} \label{eq:obj_func}
        \pi^* = \argmax_{\pi \in \Pi} \ E_{\pi} \left[ \sum_{t=0}^{\infty} \xi^{t} R_t \big| \mathcal{H}_{t-1} \right],
\end{equation}
where $\xi \in (0, 1)$ is the discount factor. It determines how far-sighted the policy will be. When $\xi$ is small, future rates weigh less on the objective function. Thus, the policy becomes myopic and focuses on maximizing the expected immediate rate $R_t$. As $\xi$ approaches 1, the policy becomes more and more farsighted. The large discount factor encourages the policy to maximize the long-term rate over multiple time slots. Therefore, the policy has to account for the effects of mobility as well.

\section{Solution Approach} \label{sec:solution}

\subsection{POMDP Framework}
The problem formulation in (\ref{eq:obj_func}) aligns with the Partially Observable Markov Decision Process (POMDP) framework~\cite{kaelbling_pomdp_tutorial}. 
A POMDP is an extension of a Markov Decision Process (MDP) to systems whose underlying states cannot be observed directly. In our problem, the mobility model (\ref{eq:mob_model}) imposes a Markovian structure on the \textit{sequences} of user locations. The BS receives partial information regarding the user location through SNR and aims to maximize the  sum of discounted expected rate by choosing a beam direction and a channel index. Therefore, we pose (\ref{eq:obj_func}) as a POMDP. Formally, a POMDP is specified by a tuple $\langle \mathcal{S}, \mathcal{A}, T, \bar{R} , \mathcal{Z}, O \rangle$, where $\mathcal{S}$ is the state space and $\mathcal{A}$ is the action space. $T(s, a, s') = P(s' | s, a)$, $s,s' \in \mathcal{S}$, $a \in \mathcal{A}$ is called the transition function. It gives the probability of transitioning from state $s$ to $s'$ after taking action $a$. $\bar{R}(s,a), \ s \in \mathcal{S}, \ a \in \mathcal{A}$ is the reward function. It quantifies the immediate reward obtained for taking action $a$ in state $s$. $\mathcal{Z}$ is the set of observations and $O(s,a,z) = P(z | s,a), \ s \in \mathcal{S}, \ a \in \mathcal{A}, \ z \in \mathcal{Z}$ is the observation function, which gives the probability of observing $z$ by taking action $a$ in state $s$~\cite{kaelbling_pomdp_tutorial}.

The state space $\mathcal{S}$ for this problem is the sequences of user locations of length $w$ i.e., $s = (u_1, \ldots, u_w) \in \mathcal{U}^{w}$. The action space $\mathcal{A}$ consists of the beam direction and channel index tuples, $a = (\hat{\theta}^a, \hat{\phi}^a, q^a) \in \Theta \times \Phi \times \mathbb{N}_{\leq C}$. The transition function is determined by the mobility model (\ref{eq:mob_model}) as
\begin{equation} \label{eq:trans_prob}
    \begin{aligned}
        T(s,a,s') &= P(s'| s, a) \quad s,s' \in \mathcal{S}, \ a \in \mathcal{A} \\
        &= P(u_1', \ldots, u_w' | u_1, \ldots, u_w) \\
        &= P\mob(u_w' , \mathbf{u}^w) \mathds{1}(u_1' = u_2, \ldots, u_{w-1}' = u_w).
    \end{aligned}
\end{equation}
The last equality comes from the definition of states and the mobility model in (\ref{eq:mob_model}). Indicator function ensures that the physical constraints resulting from the state description are satisfied. 
Since the states $s,s' \in \mathcal{S}$ are defined as $w$-tuples consisting of consecutive user locations, the probability of transition from state $s$ to $s'$ could be non-zero only if the first $(w-1)$ elements of $s'$ agree with the last $(w-1)$ terms of $s$.

According to the problem formulation in (\ref{eq:obj_func}), the reward function in this setup is the rate. By combining (\ref{eq:rec_sig}), (\ref{eq:SNR}) and (\ref{eq:rate}), it can be expressed as a function of the state and action as:
\begin{equation} \label{eq:reward_func}
    \begin{aligned}
        \bar{R}(s,a) &= \bar{R}(u_1,\ldots,u_w, \hat{\theta}^a, \hat{\phi}^a, q^a) \\
        &= R \left(r^{u_w}, \theta^{u_w}, \phi^{u_w}, \hat{\theta}^a, \hat{\phi}^a, q^a \right).
    \end{aligned}
\end{equation}
Note that the channel index $q^a$ automatically determines the frequency, bandwidth and number of antennas in (\ref{eq:rec_sig}). Furthermore, (\ref{eq:rec_sig}) indicates that the instantaneous rate in (\ref{eq:rate}) depends only on the current location, $u_w$, of the user. Since (\ref{eq:rate}) depends on SNR and SNR is a function of noise (\ref{eq:SNR}), the rewards are stochastic.
For simplicity, the expected rewards for action $a \in \mathcal{A}$ in different states $s \in \mathcal{S}$ are collected in a vector as
\begin{equation} \label{eq:exp_reward_vec}
    \mathbf{\bar{r}}_a \coloneq \Big[ E\big[\bar{R}(s_1,a) \big], \ldots, E\big[\bar{R}(s_{|\mathcal{S}|},a)\big] \Big]^T \in \mathbb{R}^{|\mathcal{S}|}.
\end{equation}

The observation space of this problem comprises of SNR measurements. Let $G(r, \theta, \phi, \hat{\theta}, \hat{\phi}, q)$ be defined as
\begin{footnotesize}
\begin{equation}\label{eq:gain_fonk}
    \begin{aligned}
        G\left(r, \theta, \phi, \hat{\theta}, \hat{\phi}, q \right) &\coloneq \frac{K P_T}{r^{\eta} f_{q}^2 N_y^q N_z^q} \times \\
        & \left[\frac{\sin \left(\pi N_y^{q} \left(\psi_{q} - \hat{\psi}_{q} \right) \right)}{\sin\left(\pi \left(\psi_{q} - \hat{\psi}_{q} \right) \right)}
        \frac{\sin\left(\pi N_z^{q} \left(\zeta_{q} - \hat{\zeta}_{q} \right) \right)}{\sin \left(\pi \left(\zeta_{q} - \hat{\zeta}_{q} \right) \right)} \right]^2,
    \end{aligned}
\end{equation}
\end{footnotesize}
where $\psi_q$, $\zeta_q$, $\hat{\psi}_q$ and $\hat{\zeta}_q$ are the normalized spatial angles in (\ref{eq:norm_spatial_angs}) for the physical angles $\theta$, $\phi$, $\hat{\theta}$ and $\hat{\phi}$, respectively. Then, the conditional probability density function (pdf) of $\gamma$ given $G(r, \theta, \phi, \hat{\theta}, \hat{\phi}, q)$ can be calculated from (\ref{eq:SNR}) as
\begin{footnotesize}
\begin{equation} \label{eq:SNR_pdf}
    g_{\gamma} \left(x | r, \theta, \phi, \hat{\theta}, \hat{\phi}, q \right) = \frac{G\left(r, \theta, \phi, \hat{\theta}, \hat{\phi}, q\right)}{\sigma_{q}^2 \ x^2} \exp \left(- \frac{G\left(r, \theta, \phi, \hat{\theta}, \hat{\phi}, q \right)}{\sigma_{q}^2 \ x} \right).
\end{equation}
\end{footnotesize}
The challenge in (\ref{eq:SNR_pdf}) is that $\gamma$ is a continuous random variable. While continuous spaces are addressed in some works, e.g.,~\cite{poupart-pomdp-obs-agg,poupart-cont-pomdp-pbvi}, the resulting solution methods are computationally complex. Therefore, most POMDP formulations are solved after discretization of action and observation spaces~\cite{poupart-cont-pomdp-pbvi}. Similarly, we discretize (\ref{eq:SNR_pdf}) in this work. 
Given some upper and lower limits $\gamma\mx$ and $\gamma\mn$, we take $M_z-1$ points $\{\gamma^1, \ldots, \gamma^{M_z-1} \}$ from the $\left[ \gamma\mn, \gamma\mx \right]$ interval and map the SNR values that are in the same interval $[\gamma^{i-1}, \gamma^{i})$ to discrete observation $z_i$. Formally, $z_i \coloneq \{\gamma \ | \ \gamma \in [\gamma^{i-1} , \gamma^{i})\} \ \forall i \in \mathbb{N}_{\leq M_z}$. Then
\begin{equation} \label{eq:SNR-discretization}
    P\left(z_i | r, \theta, \phi, \hat{\theta}, \hat{\phi}, q \right) = P\left(\gamma^{i-1} \leq \gamma < \gamma^{i} | r, \theta, \phi, \hat{\theta}, \hat{\phi}, q \right).
\end{equation}
$\gamma^0$ and $\gamma^{Mz}$ are set to $0$ and $\infty$, respectively. The resulting discretized observation set is $\mathcal{Z} =\{z_1, \ldots, z_{M_z} \}$ with $|\mathcal{Z}| = M_z$, and the observation function is given by
\begin{equation}
    \begin{aligned}
        O(s, a, z) \coloneq P(z | s,a) = P(z| r^{u_w}, \theta^{u_w}, \phi^{u_w}, \hat{\theta}^a, \hat{\phi}^a, q^a).
    \end{aligned}
\end{equation}


\subsection{Belief States}
Since the agent is not aware of the true state in a POMDP, it keeps a probability distribution called \textit{belief} over the states and makes decisions accordingly. The definition of the belief at time $t$ is 
\begin{equation} \label{eq:belief_def}
    \mathbf{b}_t = \left[ P(s_t=s_1 | \mathcal{H}_{t-1}), \ldots , P(s_t = s_{|\mathcal{S}|} | \mathcal{H}_{t-1} ) \right]^T.
\end{equation}

The belief vector is updated after every observation according to belief update rule given by
\begin{equation} \label{eq:belief_update}
    \mathbf{b}_{t+1}[s'] = \frac{\sum_{s} T(s,a,s') O(s,a,z) \mathbf{b}_t[s]}{\sum_{\tilde{s}} O(\tilde{s}, a ,z) \mathbf{b}_t[\tilde{s}]}.
\end{equation}
(\ref{eq:belief_update}) is derived from (\ref{eq:belief_def}) using the Bayes' Theorem.

Notice that (\ref{eq:belief_def}) maps $\mathcal{H}_{t-1}$ to a probability distribution over $\mathcal{S}$. In fact,~\cite{astrom_pomdp} shows that $\mathbf{b}_t$ encodes all the information in $\mathcal{H}_{t-1}$. Therefore, instead of the history, the agent can work with beliefs to make decisions. Then, (\ref{eq:obj_func}) reduces to finding the optimal mapping  $\pi(\cdot)$ from  belief vectors to actions. Let the probability simplex over $\mathcal{S}$ be $\Delta(\mathcal{S})$. Then, the optimal policy is a function $\pi: \Delta(\mathcal{S}) \rightarrow \mathcal{A}$ such that
\begin{equation} \label{eq:obj_func2}
    \begin{aligned}
        \pi^* = \argmax_{\pi \in \Pi} \ E_{\pi} \left[ \sum_{t=0}^{\infty} \xi^{t} R_t \big| \mathbf{b}_0\right],
    \end{aligned}
\end{equation}
where $\mathbf{b}_0$ is the initial belief vector and $\Pi$ is the set of all policies.

\subsection{PBVI Algorithm}

(\ref{eq:obj_func2}) is generally solved via Value Iteration Algorithm. Value function of a belief state $\mathbf{b}$ under policy $\pi(\cdot)$ is defined as the expected cumulative reward to be accumulated starting from belief $\mathbf{b}$ and following policy $\pi(\cdot)$~\cite{cassandra_witness}, i.e.,
\begin{equation} \label{eq:val_func_def}
    V_{\pi} (\mathbf{b}) \coloneq E_{\pi} \left[ \sum_{t=0}^{\infty} \xi^t R_{t} \big| \mathbf{b}_0 = \mathbf{b} \right].
\end{equation}
By definition (\ref{eq:obj_func2}), the optimal policy $\pi^*(\cdot)$ achieves the maximum value over the entire belief space. Thus, finding $\pi^*(\cdot)$ is equivalent to finding $V_*$, the value function of the optimal policy. 


The posterior $\mathbf{b}^{a,z}$ of any belief vector $\mathbf{b}$ after taking action $a$ and observing $z$ can be calculated using (\ref{eq:belief_update}). Hence, the value function $V_*$ can be expressed recursively as~\cite{pineau-point-based-pomdp-survey}:
\begin{equation} \label{eq:val_func_recur}
    V_*(\mathbf{b}) = \max_{a \in \mathcal{A}} \  \mathbf{\bar{r}}_a^T \mathbf{b} + \xi \sum_{z} P(z | b, a) V_*(\mathbf{b}^{a,z}).
\end{equation}
(\ref{eq:val_func_recur}) defines a system of linear equations, but since the beliefs are continuous, it must be solved using a different approach. $V_*$ is shown to be piecewise linear and convex in beliefs, i.e., it is the upper surface of a set of hyperplanes~\cite{kaelbling_pomdp_tutorial}. Therefore, it can be expressed using a set of vectors called $\alpha$-vectors. Each $\alpha$-vector corresponds to the value function of a policy tree, hence has an associated action $a(\boldsymbol{\alpha})$~\cite{littman_witness}. Given the set of $\alpha$-vectors $\mathcal{V}^*$ of the optimal policy, the optimal action at any belief point $\mathbf{b}$ is simply to perform the action associated with the $\alpha$-vector which has the maximum inner product with $\mathbf{b}$~\cite{kaelbling_pomdp_tutorial}, i.e. $\pi^*(\mathbf{b}) = a(\boldsymbol{\alpha^*(\mathbf{b})})$ where
\begin{equation}
    \begin{aligned}
        V_{*}(\mathbf{b}) &= \max_{\alpha \in \mathcal{V}^*} \ \boldsymbol{\alpha}^T \mathbf{b} \quad \mathrm{and} \quad \boldsymbol{\alpha}^*(\mathbf{b}) = \argmax_{\alpha \in \mathcal{V}^*} \boldsymbol{\alpha}^T \mathbf{b}.\\
    \end{aligned}
\end{equation}
However, finding the $\alpha$-vectors is generally a difficult task. It requires constructing the optimal finite horizon policies and expanding the horizon until the value function converges~\cite{cassandra_witness}. The set of $\alpha$-vectors can grow quickly in many problems, which makes finding the optimal solution challenging. Therefore, instead of solving (\ref{eq:val_func_recur}) exactly, we find an approximation to it using Point-Based Value Iteration (PBVI) algorithm~\cite{pineau_pbvi}.


The main idea behind the PBVI algorithm~\cite{pineau_pbvi} is to solve (\ref{eq:val_func_recur}) over a finite set of belief points $\mathcal{B} = \{\mathbf{b}_1, \ldots, \mathbf{b}_q \}$, then to slowly expand the belief set. It starts with the initial belief $\mathcal{B} = \{b_0\}$ and a very loose lower bound $\boldsymbol{\alpha}_0[s] = (1-\xi)^{-1} \min_{s',a} E[\bar{R}(s',a)] $, $\forall s \in \mathcal{S}$  on $V_*$. It generates an approximation to $V_*$ by interleaving the belief expansion and point-based backup stages. 


In the point-based value backup stage the value function is updated over a fixed set of beliefs $\mathcal{B}$. Given a set $\mathcal{V}_{\mathcal{B}}^j$, a new set $\mathcal{V}_{\mathcal{B}}^{j+1}$ of $\alpha$-vectors is generated by treating (\ref{eq:val_func_recur}) as an iteration. The backup operation in (\ref{eq:backup_complete}) is performed for all belief vectors in $\mathcal{B}$. For any belief vector $\mathbf{b} \in \mathcal{B}$,
\begin{equation} \label{eq:backup_complete}
    \begin{aligned}
        V_{\mathcal{B}}^{j+1}(\mathbf{b}) &= \max_{a \in \mathcal{A}} \mathbf{\bar{r}_a}^T \mathbf{b} + \xi \sum_{z} P(z | \mathbf{b}, a) \max_{\boldsymbol{\alpha} \in \mathcal{V_{\mathcal{B}}}^{j}} \boldsymbol{\alpha}^T \mathbf{b}^{a,z} \\
         &= \max_{a \in \mathcal{A}} \mathbf{\bar{r}_a}^T \mathbf{b} + \xi \sum_{z} \max_{\boldsymbol{\alpha} \in \mathcal{V_{\mathcal{B}}}^{j}} \boldsymbol{\alpha}_{a,z}^T\mathbf{b} \\
         &= \max_{a \in \mathcal{A}} \boldsymbol{\alpha}_{a,\mathbf{b}}^T \mathbf{b},
    \end{aligned}
\end{equation}
where $\boldsymbol{\alpha}_{a,z}[s] = \sum_{s'} O(s,a,z) T(s,a,s') \boldsymbol{\alpha}[s']$, $\forall \boldsymbol{\alpha} \in \mathcal{V}_{\mathcal{B}}^j$ are the projection vectors and $\boldsymbol{\alpha}_{a, \mathbf{b}} = \mathbf{\bar{r}_a} + \xi \sum_{z} \max_{\boldsymbol{\alpha} \in \mathcal{V_{\mathcal{B}}}^{j}} \boldsymbol{\alpha}_{a,z}^T\mathbf{b}$. At the beginning of every backup stage, the projection vectors $\boldsymbol{\alpha}_{a,z}$, $\forall a \in \mathcal{A}$, $\forall z \in \mathcal{Z}$ are generated and stored. Then, backups are performed for every belief point in the set $\mathcal{B}$ according to (\ref{eq:backup_complete}). This process is repeated until $V_{\mathcal{B}}$ converges. After that, the algorithm moves onto the belief expansion stage.

Once the value function converges over a belief set $\mathcal{B}$, the set must be expanded for further improvement. In doing so, PBVI focuses on beliefs that are more likely to be reached from the current set, and aims to span this reachable space as densely as possible~\cite{pineau_pbvi}. Given a distance metric $\mathcal{L}$, the distance of a belief vector $\mathbf{b}'$ to set $\mathcal{B}$ is defined as
\begin{equation} \label{eq:dist_to_set}
    \big| \mathbf{b}' - \mathcal{B} \big|_{\mathcal{L}} \coloneq \min_{\mathbf{b} \in \mathcal{B}} \big| \mathbf{b}' - \mathbf{b} \big|_{\mathcal{L}}.
\end{equation}
(\ref{eq:dist_to_set}) indicates that the distance of a belief vector $\mathbf{b}'$ to set $\mathcal{B}$ is the minimum distance between $\mathbf{b}'$ to any vector in $\mathcal{B}$. During the expansion stage, PBVI adds a successor $\mathbf{b}'$ to the set for each belief $\mathbf{b} \in \mathcal{B}$. First, a set of candidate successors $\tilde{B} = \{ \mathbf{b}^{a_1}, \ldots, \mathbf{b}^{a_{|\mathcal{A}|}} \}$ is generated by forward simulation. Then, the candidate that is the most distant from $\mathcal{B}$ is selected as the successor,
\begin{equation} \label{eq:bel_successor}
    \mathbf{b}' = \argmax_{\mathbf{b}^{a} \in \tilde{B}} \big| \mathbf{b}^a - \mathcal{B} \big|_{\mathcal{L}} \quad \forall \mathbf{b} \in \mathcal{B}.
\end{equation}
The goal in (\ref{eq:bel_successor}) is to select the new belief points judiciously such that the belief set spans the reachable space evenly. Since a new vector is generated from each belief in the set, the size of $\mathcal{B}$ doubles after each expansion in the worst case.

PBVI is an anytime algorithm. Although it can converge to the optimal value function $V_*$ in some specific cases, it usually provides an approximation. Let $V_{\mathcal{B}}$ be the PBVI approximation to the optimal value function $V_*$ over the set $\mathcal{B}$ and denote the set of corresponding $\alpha$-vectors by $\mathcal{V}_{\mathcal{B}}$. Then, the policy obtained through PBVI is given by
\begin{equation} \label{eq:pol_PBVI}
    \begin{aligned}
        \pi_{\mathcal{B}}(\mathbf{b}) = a \left(\argmax_{\boldsymbol{\alpha} \in \mathcal{V}_{\mathcal{B}}} \boldsymbol{\alpha}^T\mathbf{b} \right),
    \end{aligned}
\end{equation}
where $a(\boldsymbol{\alpha})$ is defined as the action determined by $\boldsymbol{\alpha}$.

Solution of the PBVI algorithm is usually suboptimal, i.e., the policy in (\ref{eq:pol_PBVI}) is not equivalent to the optimal policy in (\ref{eq:obj_func2}) in general. However, its performance with respect to $\pi^*(\cdot)$ can be quantified. It was shown in~\cite{pineau_pbvi} that the error $|| V_* - V_{\mathcal{B}}||_{\infty}$ is bounded by how densely $\mathcal{B}$ samples the belief simplex $\Delta(\mathcal{S})$. Therefore, the approximation error usually becomes small once the belief set gets sufficiently large. 



\section{Evaluations} \label{sec:numeric}

In this section, the performance and parameter dependence of solutions generated by the PBVI policy are evaluated. First, the average throughput performance of the policy has been scrutinized as a function of a critical parameter related to the uncertainty in user mobility. Then, we assess the robustness of the policy to deviations of the user mobility from the assumed mobility model.

\subsection{System Setup and Parameters}
We consider a 3D coordinate system as shown in Fig.~\ref{fig:UPA}. The BS is located at the origin and its height is $h_t=10$m, which places the ground at $z=-10$ plane. The road is parallel to the $y$-axis and its distance to the BS's projection on the ground is $10$m. The total length of the road segment is $240$m. The BS is at the midpoint of the road, so the road forms a straight line between the points $(10, -120, -10)$ and $(10, 120, -10)$ in Cartesian coordinates. Assuming the height of the UE is $h_r=1.5$m from the ground, the UE travels along a straight line between $(10, -120, -8.5)$ and $(10, 120, -8.5)$, which is divided into $M_r=12$ cells. Each cell $u_i$ for $i \in \mathbb{N}_{\leq M_r}$ is represented by the spherical coordinates $(r^{u_i}, \theta^{u_i}, \phi^{u_i})$ of its center. Note that $\theta^{u_i} \in [-\pi/2, \pi/2]$ and $\phi^{u_i} \in [-\pi/2, 0]$ $\forall i \in \mathbb{N}_{\leq M_r}$.

\begin{table}[!t]
\renewcommand{\arraystretch}{1.3}
\caption{Mobility Model $P\mob(u_t, u_{t-1}, u_{t-2})$}
\label{tab:trans_prob}
\centering
\begin{tabular}{|c|c|c|c|}
\hline
\bfseries $u_{t}$ & \bfseries $u_{t-1}$ & \bfseries $u_{t-2}$ & \bfseries $P(u_{t}|u_{t-1}, u_{t-2})$\\
\hline\hline
$u_{i}$ & $u_{i}$ & $u_{M_r+1}$ & $p$\\
\hline
$u_{i \pm 1}$ & $u_{i}$ & $u_{M_r+1}$ & $0.5(1-p)$\\
\hline
$u_{i}$ & $u_{i}$ & $u_{i}$ & $\kappa_2 p$\\
\hline
$u_{i \pm 1}$ & $u_{i}$ & $u_{i}$ & $0.5(1-\kappa_2 p)$\\
\hline
$u_{i}$ & $u_{i}$ & $u_{i \pm 1}$ & $p$\\
\hline
$u_{i \mp 1}$ & $u_{i}$ & $u_{i\pm 1}$ & $\kappa_1(1-p)$\\
\hline
$u_{i \pm 1}$ & $u_{i}$ & $u_{i\pm 1}$ & $(1-\kappa_1)(1-p)$\\
\hline
\end{tabular}
\end{table}

\begin{figure*}[!t]
\begin{subfigure}{0.48\linewidth}
    \centering
    \includegraphics[width=\textwidth]{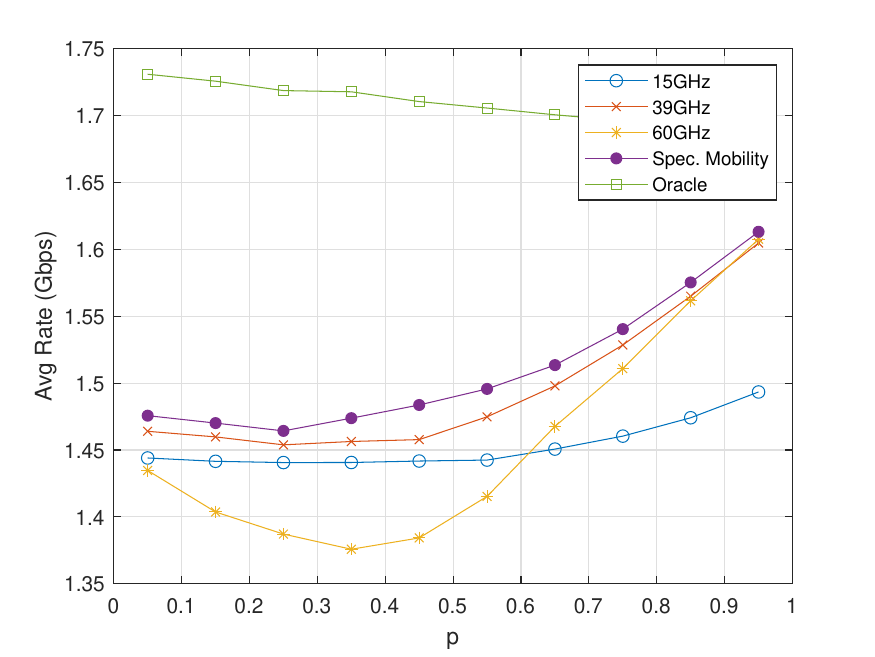}
    \caption{Average rate vs $p$}
    \label{fig:15-39_mob_model_rate}    
\end{subfigure}
\hfill
\begin{subfigure}{0.48\linewidth}
    \centering
    \includegraphics[width=\textwidth]{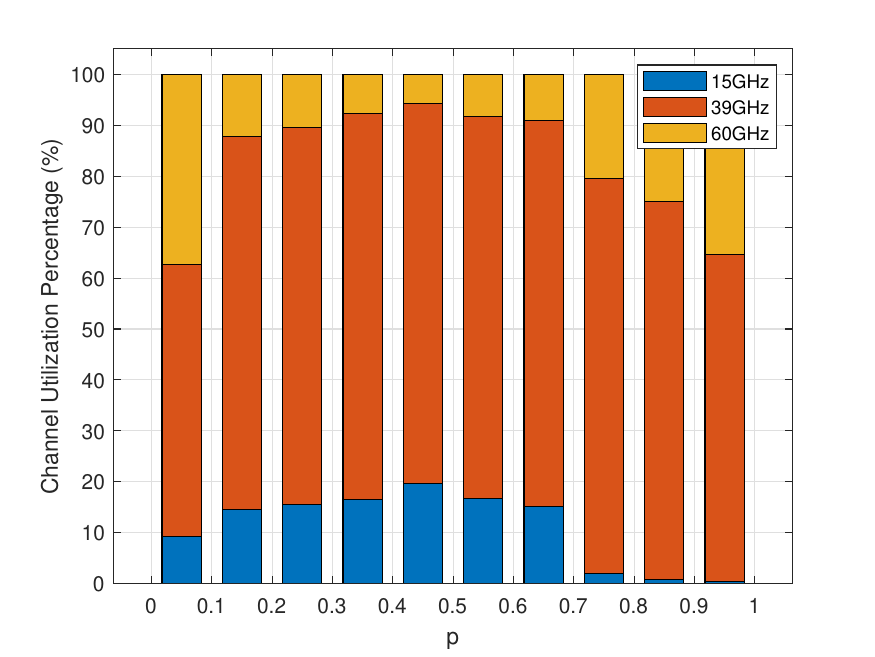}
    \caption{Channel utilization percentage by spectrum mobility vs $p$}
    \label{fig:15-39_mob_model_freq_pct}    \end{subfigure}
\caption{Performance of the Spectrum Mobility vs. mobility model parameter $p$}
\label{fig:mob-model-monte-carlo}    
\end{figure*}

We take the window size of the mobility model (\ref{eq:mob_model}) as $w=2$ and assume that, given $u_{t-1} = u_i$, the user can at most move to one of the adjacent cells in the next time slot, i.e., $u_t \in \{u_{i-1}, u_i, u_{i+1} \}$. The transition probabilities $P\mob(u_{t}, u_{t-1}, u_{t-2})$ are given in Table~\ref{tab:trans_prob}, where $u_{M_r+1}$ indicates that the user is outside the coverage area. According to this model, when there is no history of the mobility pattern, e.g. rows 1-2 of Table~\ref{tab:trans_prob}, the user stays within the same cell with probability $p$ or moves to one of the adjacent cells with probability $0.5(1-p)$. If the user has been in the same position within the last $w$-window, e.g. rows 3-4 of Table~\ref{tab:trans_prob}, the probability of staying in the same cell scales down with $\kappa_2$. Meanwhile, moving to adjacent cells is equiprobable with $0.5(1-\kappa_2p)$. Lastly, if the history indicates a mobility pattern towards a certain direction, e.g. rows 5-7 of Table~\ref{tab:trans_prob}, the probability of moving in that direction scales up with $\kappa_1$ and moving in the opposite direction scales down with $(1-\kappa_1)$. We set $\kappa_1 = \kappa_2 = 0.95$ in our simulations.

We use $\mathcal{F} = \{15, 39, 60 \}$GHz for channel frequencies. $39$GHz and $60$GHz are standard frequencies from FR2 bands \cite{3GPP-FR2}. We also included $15$GHz in our simulations because the FR3 band, to which $15$GHz belongs, is also being considered for deployment \cite{rappaport-fr1-fr3-propagation-models}. Assuming that the bandwidth allocations for $15$GHz will be similar to FR1 band \cite{3GPP-FR1}, the channel bandwidths are assigned as $\mathcal{W} = \{90, 100, 100\}$MHz, respectively. Considering a shared aperture area of $37.5\mathrm{mm} \times 37.5\mathrm{mm}$, the number of elements in the UPAs are roughly calculated as $\{4 \times 4, \ 10 \times 10, \ 16\times 16\}$. The SNR values (\ref{eq:SNR-discretization}) are discretized to 25 levels on a logarithmic scale ranging from $-50$dB to $80$dB. The path loss exponent and the total transmit power are set to $\eta = 2$ and $P_T = 1$ W, respectively. The spectral density of noise is taken to be $-174$ dBm/Hz \cite{proakis-comm-systems}. The path gain constant is determined using Frii's Transmission formula as $K = \left(c / (4 \pi) \right)^2 \approx 5.7 \times 10^{14}$.

We run the PBVI algorithm for 4 steps and perform belief expansion twice in each step. The discount factor is set to $\xi=0.99$. Instance of the PBVI algorithm that considers all three frequency bands is referred to as ``Spectrum Mobility''. Policies that use only one of the frequency bands in $\mathcal{F} = \{15, 39, 60 \}$GHz are also computed with PBVI. These are essentially single-frequency versions of the spectrum mobility and are referred to as such in the following sections.  We define an agent called oracle to quantify the extent to which uncertainty affects the performance of the spectrum mobility. Oracle knows the exact location of the user at any time and always performs the optimal action (\ref{eq:opt_sol_perf_info}).



\begin{figure*}[!t]
\begin{subfigure}{0.48\linewidth}
    \centering
    \includegraphics[width=\textwidth]{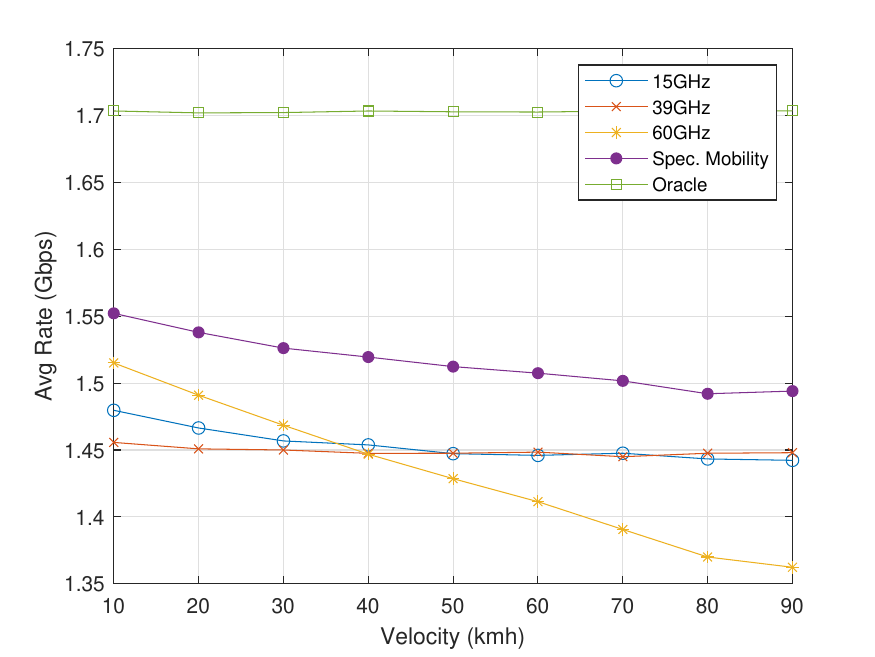}
    \caption{$p=0.35$}
    \label{fig:15-39_robustness_q35}    
\end{subfigure}
\hfill
\begin{subfigure}{0.48\linewidth}
    \centering
    \includegraphics[width=\textwidth]{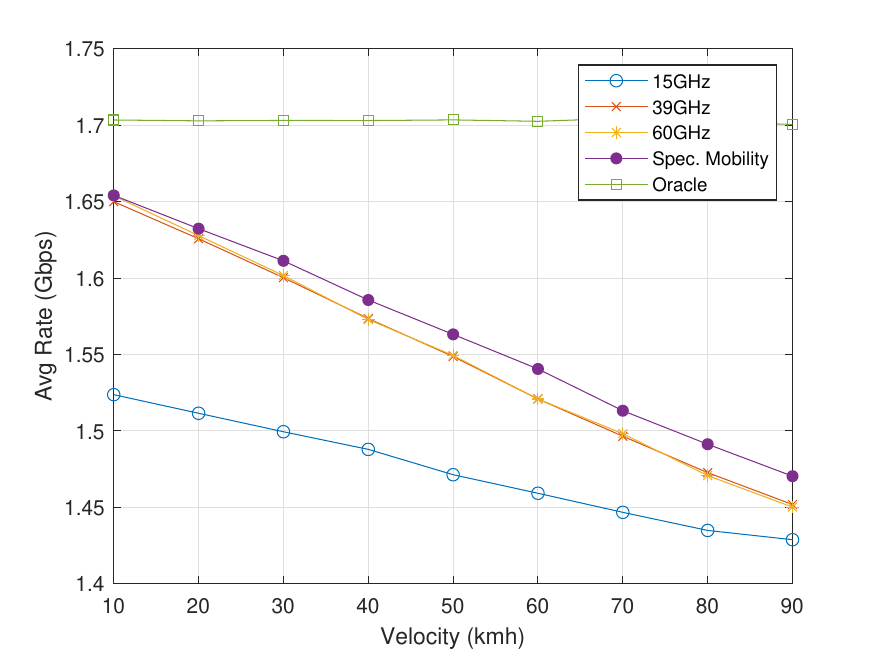}
    \caption{$p=0.95$}
    \label{fig:15-39_robustness_q95}    \end{subfigure}
\caption{Performance of the Spectrum Mobility on a fixed path for given mobility model parameters $p$}
\label{fig:mob-model-robustness}    
\end{figure*}

\subsection{Spectrum Mobility Performance}
The average data rates achieved by spectrum mobility and its single-frequency counterparts are shown in Fig.~\ref{fig:15-39_mob_model_rate} for various $p$ values. Sample paths are generated randomly according to the mobility model for each $p$ parameter. The results are obtained by averaging 5,000 Monte Carlo simulations. 

In Fig.~\ref{fig:15-39_mob_model_rate}, spectrum mobility is observed to achieve a higher average rate than its single-frequency versions for all $p$ values. This is explained by the adaptability of the spectrum mobility to uncertainty. The only decision parameter in single-frequency variants of spectrum mobility is the beam direction. The channel frequency, antenna array and bandwidth are fixed. Meanwhile, spectrum mobility can cruise over the spectrum to find the best combination that yields the highest data rate for different levels of uncertainty. This is also supported by the channel utilization percentages shown in Fig~\ref{fig:15-39_mob_model_freq_pct}. In high-uncertainty scenarios, which correspond to mid-range $p$ values (Table~\ref{tab:trans_prob}), utilization percentage of $15$GHz channel varies between $15-20 \%$ in our simulations. However as $p$ approaches 1, this percentage rapidly drops and becomes almost $0$. This is due to the fact that there are fewer antenna elements in the UPA at $15$GHz. Thus, the beams are wider and the beam misalignment resulting from uncertainty can be better tolerated. The propagation characteristics of $15$GHz are also more favorable. Hence, despite its lower bandwidth, $15$GHz is still utilized by the spectrum mobility in mid-range $p$ parameters. The converse is true for $60$GHz. UPA consists of more antenna elements to combat the high attenuation experienced at $60$GHz. This results in narrower beams with higher maximum gain but also increases the risk of beam misalignment. Hence, we observe a $2$ to $6$ fold increase in the utilization of $60$GHz in our simulations as $p$ values approach to either extreme, i.e. $0$ or $1$.

Among the set of channel frequencies, $39$GHz is the most widely used in this setup because it is the middle ground between high-gain-high-risk $60$GHz and the low-gain-low-risk $15$GHz. As shown in Fig.~\ref{fig:15-39_mob_model_rate}, it achieves a $3\%$ higher data rate on average compared to other single-frequency versions of spectrum mobility in our simulations. The main reason why it outperforms $60$GHz is its favorable propagation characteristics and smaller array size. Given perfect location information (\ref{eq:opt_sol_perf_info}), the average rate of $60$GHz channel would be about $11\%$ higher than that of $39$GHz. However, the uncertainty stemming from mobility significantly degrades the performance of $60$GHz as the risk of beam misalignment increases and this makes $39$GHz more favorable. On the other hand, the main reason why $39$GHz outperforms $15$GHz is its bandwidth. On average, $15$GHz achieves 
$2.8\%$ less data rate than $39$GHz in our simulations. However, if the $15$GHz and $39$GHz channels had the same bandwidth, the former would have achieved about $11\%$ higher data rates given perfect location information (\ref{eq:opt_sol_perf_info}). Considering that $15$GHz has better propagation characteristics and wider beams, its performance would stand out even further under mobility. 

Although it may look like $39$GHz provides a reasonably close approximation of spectrum mobility, this significantly depends on system parameters. The algorithm decisions are highly dependent on and sensitive to system parameters, including frequency bands, the number of antennas, bandwidths, road structure, and mobility model. Regardless of other system parameters, having multiple frequency bands at disposal remains an advantage as it enhances flexibility and robustness. 

\subsection{Robustness Study}
Robustness of the spectrum mobility to deviations from mobility model is demonstrated in Fig.~\ref{fig:mob-model-robustness}. The user travels on a fixed path stretching from one end of the road to the other at a constant speed. Assuming that each time slot is $0.25$s \cite{dimitros-operational-mmwave-networks}, the trajectory is simulated 5,000 times for different velocities and the resulting average data rate is shown in  Fig.~\ref{fig:mob-model-robustness} for two different mobility model parameters. We select $p=0.35$ and $p=0.95$ in our simulations for multiple reasons. Firstly, when the paths were generated according to the mobility model in the previous section, the minimum and maximum rates across all policies, including the spectrum mobility and its single-frequency variants, occur at $p=0.35$ and $p=0.95$, respectively. Secondly, since $p=0.35$ corresponds to relatively high-uncertainty scenarios, it can be used for a broad set of trajectories without compromising much from the performance, i.e., it is more generalizable. In contrast, the mobility model matches the trajectory quite well for low speeds when $p=0.95$ but the same cannot be said for high speeds. Therefore, we chose these two contrasting parameter values to examine the generalization properties of the spectrum mobility.

Spectrum mobility is shown to achieve higher data rates than its single-frequency variants in both Fig.~\ref{fig:15-39_robustness_q35} and Fig.~\ref{fig:15-39_robustness_q95} over a speed range of $10-90$kmh. It achieves an average increase of $4.8\%$ and $2.45\%$ in rate in our simulations for $p=0.35$ and $p=0.95$, respectively. 

A general comparison between $p=0.35$ and $p=0.95$ reveals that the rates of all policies are $1.3-8.2\%$ higher on average when $p=0.95$ but they are also more sensitive to speed variations. As the velocity varies between $10-90$kmh the rates of all policies decrease by $6.4-13.1\%$. Meanwhile, this ratio is between $0.74-10.7\%$ when $p=0.35$. Specifically for spectrum mobility, the rate drops by $4\%$ when $p=0.35$ while it decreases by $11.8\%$ for $p=0.95$. 

These results show that the spectrum mobility is efficient even when the user trajectory deviates from the mobility model.


\section{Conclusions and Future Work} \label{sec:conclusions}

In this work, we investigate the spectrum mobility and antenna management problem for efficient beamforming under mobility. The BS can access different parts of the spectrum and aims to increase the expected data rate by selecting the best channel frequency and beam direction. Different frequency bands have different bandwidths, number of antennas, and propagation characteristics. When this is combined with the uncertainty stemming from mobility, a natural trade-off arises between maintaining the link connection and achieving high data rates. Thus, the problem becomes finding a resource allocation policy that achieves high data rates under varying levels of uncertainty. We formulate this problem as a POMDP and use PBVI algorithm to find an approximate solution. The resulting policy can be precomputed and stored for online access. Simulations show that adopting the channel frequency and beamwidth according to uncertainty in the user location can increase the average data rates and improve the robustness of beam tracking. 

This work establishes a foundation for multifaceted beam management schemes that involve both frequency and the number of antenna elements as resources. In our future work, we will consider systems where the parameters of the mobility model are unknown and must be learned from previous actions and observations. We will explore adaptive solutions for changing system dynamics. Moreover, we plan to generalize this work to a multi-user setup. We will also investigate different solution approaches, such as reinforcement learning, that can yield faster and online algorithms.


\bibliographystyle{IEEEtran}{}
\bibliography{References}

\end{document}